\documentclass[12pt,preprint]{aastex}
\newcommand{\mkrc}{\ensuremath{ {M}_{K}^{RC} }}
\newcommand{\krc}{\ensuremath{ {K}_{RC} }}
\newcommand{\ak}{\ensuremath{ {A}_{K} }}
\newcommand{\mmo}{\ensuremath{ {(m-M)}_{0} }}

\newcommand{\ebv}{\ensuremath{E(B-V)}}
\def\lea{\mathrel{<\kern-1.0em\lower0.9ex\hbox{$\sim$}}}
\def\gea{\mathrel{>\kern-1.0em\lower0.9ex\hbox{$\sim$}}}

\begin{document}


\title{Distances to Populous Clusters in the LMC via the K-band Luminosity
of the Red Clump}

\author{Aaron J. Grocholski\altaffilmark{1}, Ata Sarajedini}
\affil{Department of Astronomy, University of Florida, P.O. Box
112055, Gainesville, FL 32611; aaron@astro.ufl.edu, ata@astro.ufl.edu}
\altaffiltext{1}{Visiting Astronomer, Cerro Tololo Inter-American 
Observatory, National Optical Astronomy Observatory, which is operated by 
the Association of Universities for Research in Astronomy (AURA), Inc., 
under cooperative agreement with the National Science Foundation.}
\author{Knut A. G. Olsen}
\affil{Cerro Tololo Inter-American Observatory, National Optical Astronomy 
Observatory, Casilla 603, La Serena, Chile; kolsen@noao.edu}
\author{Glenn P. Tiede}
\affil{Department of Physics and Astronomy, Bowling Green State 
Univeristy, Bowling Green, OH, 43403; gptiede@bgnet.bgsu.edu}
\author{and\\ Conor L. Mancone}
\affil{Department of Astronomy, University of Florida, P.O. Box
112055, Gainesville, FL 32611; cmancone@astro.ufl.edu}

\begin{abstract}

We present results from a study of the distances and distribution of a
sample of intermediate-age clusters in the Large Magellanic Cloud.  Using
deep near-infrared photometry obtained with ISPI on the CTIO 4m, we have
measured the apparent $K$-band magnitude of the core helium burning red
clump stars in 17 LMC clusters.  We combine cluster ages and metallicities
with the work of Grocholski \& Sarajedini to predict each cluster's
absolute $K$-band red clump magnitude, and thereby calculate absolute
cluster distances.  An analysis of these data shows that the cluster
distribution is in good agreement with the thick, inclined disk geometry
of the LMC, as defined by its field stars.  We also find that the old
globular clusters follow the same distribution, suggesting that the LMC's
disk formed at about the same time as the globular clusters, $\sim$ 13 Gyr
ago.  Finally, we have used our cluster distances in conjunction with the
disk geometry to calculate the distance to the LMC center, for which we
find $\mmo = 18.40 \pm 0.04_{ran} \pm 0.08_{sys}$, or $D_0 = 47.9 \pm 0.9
\pm 1.8$ kpc. 

\end{abstract}


\section{Introduction}
\label{phot:intro}

Interactions and merger events can dominate the formation histories of
galaxies, both large and small, at high and low redshift
(\citealt{abraham1999,schweizer1999}) and the Milky Way (MW) and its
satellite galaxies are an excellent example of this.  The Large Magellanic
Cloud (LMC) is a nearby satellite galaxy that is dynamically active; it
exhibits many epochs of star formation (including current star formation)  
while also suffering from tidal interactions with the Small Magellanic
Cloud (SMC) and the MW.  Given its proximity, stellar populations in the
LMC are easily resolved, allowing us to obtain information such as ages,
chemical abundances, kinematics and distances to individual stars.  Thus,
the LMC offers us an excellent local laboratory in which to study the
effects of gravitational forces on the evolution of a satellite galaxy.

Traditionally, the LMC has been treated as a planar galaxy that, despite
its proximity, can be assumed to lie at a single distance from us.  This
is in spite of the fact that, using distances to field Cepheid variables,
\citet{caldwellcoulson1986} first showed that the disk of the LMC is
inclined with respect to the sky.  More recent studies of field stars have
confirmed this finding.  For example, \citet{vdmcioni2001} combined near
infrared photometry from the Deep Near-Infrared Southern Sky Survey
(DENIS) and the Two Micron All-Sky Survey (2MASS) to study the
distribution of field stars in the LMC out to a radius of $\sim 7\degr$.  
Using both the tip of the red giant branch (RGB) and asymptotic giant
branch as relative distance indicators, they found an $I$-band
peak-to-peak sinusoidal brightness variation of $\sim$0.25 mag that
changes as a function of position angle on the sky, with stars in the
northeast portion of the LMC brighter than stars in the southwest.  
Attributing this variation in brightness to a difference in distance, they
calculated an inclination of $i = 34\fdg7 \pm 6\fdg2$ for the disk of the
LMC (where 0$\degr$ is face on) and the line of nodes position angle (the
intersection of the plane of the galaxy with the plane of the sky) of
$\Theta = 122\fdg5 \pm 8\fdg3$.  In an approach similar to
\citet{vdmcioni2001}, \citet{olsensalyk2002} use the apparent $I$-band
magnitude of core helium burning red clump (RC) stars to explore the
structure of the disk.  Calculating relative distances for 50 fields
spread across a $6\degr \times 6\degr$ area of the LMC, they find $i =
35\fdg8 \pm 2\fdg4$, in agreement with the \citet{vdmcioni2001} result,
and $\Theta = 145\degr \pm 4\degr$. In addition to the inclination, the
LMC's geometry becomes even more complex when we consider that its disk
($v/\sigma = 2.9 \pm 0.9$) is thicker than the MW's thick disk ($v/\sigma
\approx 3.9$, \citealt{vdmetal2002}) and that the disk is flared
(\citealt{alvesnelson2000}) and also possibly warped
(\citealt{olsensalyk2002,nikolaevetal2004}) as a result of interactions
with the SMC and MW.  Even with all of the knowledge of the LMC's
structure from field star studies, the spatial distribution of populous
clusters in the LMC remains relatively unexplored.  \citet[][see also
\citealt{grocholskietal2006}]{schommeretal1992} showed that the LMC
clusters have disk-like kinetmatics, however, only recently has a planar
geometry been illustrated for the LMC cluster system
(\citealt{kerberetal2006}).


Distances to stellar populations in the LMC have been calculated using a
variety of standard candles, including the period-luminosity (P-L)
relation of Cepheid variables (e.g.,~\citealt{macrietal2006};
\citealt{gierenetal1998}), the mean absolute magnitude-metallicity
relationship for RR Lyraes (e.g.,~\citealt{walker1985}), and color
magnitude diagram (CMD) features like the tip of the RGB
(e.g.,~\citealt{cionietal2000}), RC stars
(e.g.,~\citealt{udalski2000,sarajedinietal2002}), or main sequence turn
off (MSTO; \citealt{kerberetal2006}).  One standard candle that has yet to
be fully exploited, and is geared toward studying clusters, is the
$K$-band luminosity of the RC.  In their work, \citet[hereafter
GS02]{gs02} use 2MASS $JK_S$ photometry of 14 Galactic open clusters that
possess internally consistent ages, metallicities, and MSTO fitting
distances to calibrate the absolute $K$-band magnitude of the RC ($\mkrc$)
as a function of age and metallicity.  An important result from their
study is that, while variations in the RC brightness are smaller in the
$K$-band than what is seen in the $V$- or $I$-bands, $\mkrc$ varies as a
function of both age and metallicity and, for young ages ($\lea$ 3 Gyr),
$\mkrc$ can vary by up to a magnitude.  Therefore, knowledge of the
abundances {\it and} ages of RC stars, something that can only be
unequivocally gleaned from clusters, is necessary to properly employ the
RC as a standard candle.  Since this method provides an absolute distance,
its application allows the determination of both the spatial distribution
of clusters and the distance to the LMC.

The distance to the LMC has been of considerable interest in recent years,
largely due to its use as the zeropoint for the extragalactic distance
scale.  The HST Key Project to determine $H_0$ (see
\citealt{freedmanetal2001} for final results on the project) used a sample
of Cepheid variables in the LMC, along with an adopted distance of $\mmo =
18.5 \pm 0.1$ (\citealt{madorefreedman1991}), to define the fiducial
Cepheid P-L relation.  \citet{freedmanetal2001} then used this new P-L
relation to calculate distances to a large number of galaxies, thereby
allowing the calibration of secondary standard candles (Type Ia and Type
II supernovae, Tully-Fisher relation, surface brightness fluctuations,
fundamental plane) that lie further up the extragalactic distance ladder.  
Thus, the accuracy of their value of $H_0 = 72 \pm 8$ km s$^{-1}$
Mpc$^{-1}$ is ultimately determined by the accuracy of the distance to the
LMC; it turns out that the distance error constitutes 6.5\% of their 9\%
error budget.  Their adopted distance, however, was based on previously
published distances and, until recently, there have been rather large
discrepencies between different methods and sometimes even among distances
calculated using the same method (particularly with optical photometry of
the RC).  In general, the LMC distances can be split up into a ``long"  
distance of $\sim$18.5-18.7 mag, usually found with Population I
indicators, and a ``short" distance of $\sim$18.3 mag, calculated
primarily from RR Lyrae variables.  \citet{clementinietal2003} review the
LMC distances and methods in detail and find that the long and short
distance scales can be reconciled, at least to within the errors, with
improved photometry and/or reddening estimates.  From the distances they
have collected (and corrected), \citet{clementinietal2003} find a mean LMC
distance of $\mmo = 18.515 \pm 0.085$, in good agreement with the value
adopted by \citet{freedmanetal2001}.

In an effort to determine the spatial distribution of the LMC cluster
system and improve the accuracy of the distance to the LMC, we apply the
approach of GS02 to calculating absolute distances to 17 
populous
clusters in the LMC.  Cluster distances, combined with the geometry of
the cluster system allow us to determine an accurate distance to the
center of the LMC.  In \S \ref{phot:sec:data} we discuss the near-infrared
data acquisition, reduction, and photometry.  The cluster ages and
abundances necessary for accurately determining $\mkrc$ are presented in
\S \ref{phot:sec:aa} and in \S \ref{phot:sec:mags} we calculate $\krc$ and
$\mkrc$ for our cluster sample.  Finally, in \S \ref{phot:sec:moduli},
cluster distances and the distance to the center of the LMC are given,
with a comparison to selected previous works in \S \ref{phot:sec:dist}.  
Our results are summarized in \S \ref{phot:sec:summary}.

\section{Data}
\label{phot:sec:data}
\subsection{Observations}
\label{phot:sec:obs}

We have obtained near infrared images of a sample of populous LMC clusters
over the course of six nights (20-22 January 2003 and 06-08 February 2004)
at the Cerro Tololo Inter-America Observatory Blanco 4m telescope.  All
data were taken with the Infrared Side Port Imager (ISPI), which utilizes
a 2048 $\times$ 2048 HAWAII 2 HgCdTe array.  In the f/8 configuration,
ISPI has a field of view of $\sim10\arcmin \times 10\arcmin$ with a plate
scale of $\sim0\farcs 33$ pixel$^{-1}$.  At the time of our observations,
ISPI was equipped with $J$ (1.25 $\mu$m), $H$ (1.64 $\mu$m), and $K'$
(2.12 $\mu$m) filters on loan from Gemini and all clusters were imaged in
the $J$- and $K'$-bands with about half of the clusters also having
$H$-band data.  Average seeing for all six nights was $\sim1.2\arcsec$.

Each cluster was observed with a nine-point dither pattern, centered on
the cluster, with dither offsets ranging between $30\arcsec$ and
$120\arcsec$, depending on the size and density of the target.  Total
exposure time in each band was as follows: $J$ - 540s;  $H$ - 846s; $K'$ -
846s.  For the first run, $H$- and $K'$-band images were split up into
shorter exposures to ameliorate the effects of sky brightness in the
near-infrared.  As we were the first science users of ISPI, a better
understanding of the instrument, along with changes in the electronics
between observing runs, resulted in our group adjusting the exposure time
splits for the second observing run.  Specifically, due to the range over
which the ISPI detector is linear, we discovered the need to split up the
$J$-band images into shorter exposures in order to keep many of the stars
from falling into the non-linear regime.  In addition, for all three
bands, short exposures (4s at each dither point) were needed to avoid
saturating the brightest stars in the frame.  In Table
\ref{tab:exposure_times}, we detail the exposures times for each band and
observing run and in Table \ref{tab:cluster_info} we list our target
clusters along with their positions on the sky, the filters in which they
were observed, and the run during which each cluster was imaged.  For all
but one of the clusters observed during both runs, only the short (4s)
exposures were taken during the second run; the exception to this is NGC
2155, for which the entire set of $K'$ exposures was obtained during the
second run.  \subsection{Reduction}

We have processed our data using standard data reduction techniques.  All
images have been dark subtracted, sky subtracted and then flat fielded
using on-off dome flats.  For each target, sky frames were created by
median combining the dithered cluster images, thus eliminating the stars
and leaving only the sky in the final combined sky frame.  Before shifting
and combining our cluster images we had to address the problem of
geometric distortions.  ISPI's large field of view causes images to be
curved at the focal plane and, if not corrected, final frames created by
shifting and combining the dithered images will have severely degraded
image quality across much of the frame.  This problem was exacerbated by
the large offsets in our dither pattern.  Using Galactic bulge star data
kindly provided by A. Stephens (2003, private communication), we created
and applied a high order distortion correction to our images using the
IRAF tasks {\it geomap} and {\it geotran}.  Corrected images were then
aligned, shifted, and average combined and bad pixels were masked to
create a final science image for each cluster and filter.  The final image
quality was excellent and only stars near the corners of the frame
exhibited any signs of distortion.  We note that for each cluster, we have
created two science images in each band; a short exposure image, created
by combining only the 4s exposures from each dither point, and a long
exposure that is a combination of all data for a given cluster.  As
mentioned in \S \ref{phot:sec:obs}, the short exposures were necessary for
accurate photometry of the bright RGB stars.  In Fig.~\ref{fig:kbandphot},
we present $K'$-band images of an $\sim 4\arcmin \times 4\arcmin$ region
around each of our target clusters.  We have used the final combined long
exposure image for each cluster

\subsection{Photometry}

Using a combination of DAOPHOT and ALLSTAR (\citealt{stetson1987}), we
have photometered our images with the following method.  A rough PSF was
created from the brightest $\sim$200 stars in each image; we have made
sure to only choose stars that were in the linear regime of the detector.  
This rough PSF was then used to remove neighbors from around the full set
of $\sim$50-150 PSF stars (depending on cluster), which allowed us to
create a more robust PSF from the cleaned image.  Next, ALLSTAR was used
to fit the improved PSF to all stars that were detected in the science
frames.  In an effort to detect and photometer faint stars and/or
companions, we performed a single iteration where we subtracted all stars
photometered in the first ALLSTAR pass, searched for previously undetected
stars, and then measured all of the new detections and added them to the
photometry list.  Aperture corrections, calculated for each science frame,
were then applied to the PSF photometry.  Lastly, we combined the aperture
corrected photometry lists for each filter with the requirement that a
star be detected in all available bands for it to be included in the final
combined list of instrumental magnitudes.

Finally, to calibrate the instrumental photometry for each cluster, we
began by matching stars in common between our long and short exposures,
then throwing out stars that are non-linear or saturated (are bright) in
the long exposures or have large errors (are faint) in the short
exposures.  Typically, we are left with intermediate brightness stars
covering a range of $\sim$2 mag over which we calculate the offset
necessary to bring the long exposure photometry onto the `system' of the
short exposures.
For clusters imaged over two epochs (see Table \ref{tab:cluster_info}),
we find different magnitude offsets between the long and short exposures
as compared to clusters observed during only the second observing run.  
This difference is likely due to different sky conditions during our two 
observing runs.  
  After offsetting the long exposure photometry, we
combine the long and short photometry in three pieces; the bright star
photometry is taken from only the short exposures (long exposures are
non-linear or saturated) while the faint stars come only from the offset
long exposure photometry (stars have large errors or are not detected in
the short exposures).  The intermediate brightness stars, which have good
photometry from both the long and short exposures, are averaged together
for the final catalog of each cluster.  To put our photometry onto a
standard system, we match our stars with those in the All-Sky Data Release
of the Two Micron All Sky
Survey\footnote{http://www.ipac.caltech.edu/2mass/releases/allsky}
(2MASS).  We have restricted the 2MASS selection to only those stars
possessing either aperture or PSF fitting photometry and having errors
less that 0.1 mag.  Zeropoint offsets for each band are then calculated
and applied to our photometry. In the last step of our calibration, we
follow the approach of GS02 and convert our photometry (on the 2MASS
system) to the \citet{besselbrett1988} system using the conversions
presented by \citet[][their Eqs.~A1-A4]{carpenter2001}.  This step is
necessary as it places our photometry on the same system as the
\citet{girardisalaris2001} models (see \S \ref{phot:sec:mags}).  We note
that we have not fit any color terms in our calibration due to the small
range in color ($\sim$0.5 mag) covered by the RGB in addition to the
similarity of the ISPI and 2MASS filter systems.

\section{Cluster Ages and Abundances}
\label{phot:sec:aa}

As mentioned in \S \ref{phot:intro}, GS02 showed that knowledge of a
populous cluster's age and metallicity is imperative to accurately
predicting $\mkrc$, and thus determining the cluster's distance.  This is
especially true for clusters with log(Age) $\lea$ 9.3 ($\lea$ 2 Gyr) or
[Fe/H] $\lea -0.4$, two regions of parameter space where $\mkrc$ can vary
rapidly (see Figs.~5 and 6 in GS02) and in which many LMC clusters reside.

For the cluster metallicities, we turn primarily to the recent work of
\citet{grocholskietal2006}.  In their paper, they present [Fe/H] for 28
populous LMC clusters, derived from the strong near infrared absorption
lines of the \ion{Ca}{2} triplet; all but four of the clusters in our
sample (ESO 121-03, NGC 1783, NGC 1978, and SL 896) have metallicities in
\citet{grocholskietal2006}.  Red giants in NGC 1783 were studied by
A.~A.~Cole et al.~(2007, in preparation) using the \ion{Ca}{2} triplet in
an almost identical approach to that of \citet{grocholskietal2006}, so we
adopt their metallicity ($-0.47 \pm 0.14$ dex) for this cluster.  For NGC
1978, we use the metallicity calculated by \citet{ferraroetal2006}, which
is based on high resolution spectra of 11 red giant stars.  We note that
their value of $-0.38 \pm 0.07$ dex is in good agreement with the results
of A.~A.~Cole et al.~(2007, in preparation), who find [Fe/H] = $-0.35 \pm
0.07$.  Using UVES on the VLT, \citet{hilletal2000} obtained high
resolution spectra for two giant stars in ESO 121-03 and found [Fe/H] =
$-0.91 \pm 0.16$, which we will adopt for this paper.  Finally, while the
small cluster SL 896 has no previously published spectroscopically derived
[Fe/H] available, the results of \citet{grocholskietal2006} show that the
intermediate metallicity LMC clusters have a very tight spread in
metallicity ($\sigma = 0.09$), with a mean metallicity of $-0.48$ dex.  
Thus, we adopt these values as the metallicity and error for SL 896.  
Cluster metallicities and errors are presented in columns 2 and 3 of Table
\ref{tab:cluster_ages}.

As for the ages, the most reliable way to determine cluster ages is by
comparing the predictions of theoretical isochrones to the luminosity of a
cluster's main sequence turn off.  However, no large scale database of 
main sequence fitting (MSF) ages exists for LMC clusters.  To
address this shortcoming, we have begun to compile optical photometry that
reaches below the main sequence turn off (MSTO) for a large number of LMC
clusters.  While the entire study will be presented in a future paper
(Grocholski et al.~2007, in preparation), we herein provide a brief
description of the data set and fitting method that are used to derive
cluster ages, as well as present ages for a sub-sample of clusters.  
Optical photometry was taken primarily from the literature and in column 7
of Table \ref{tab:cluster_ages}, we list the CMD sources.  In a few cases,
we have used unpublished optical images, obtained with either VLT FORS2
(NGC 1846, NGC 2203, IC 2146; see \citealt{grocholskietal2006}) or {\it
HST} WFPC2 (NGC 2193; program number GO-5475).  For the three clusters
with $V$ and $I$ band VLT FORS2 images, stars were identified and
photometered with the aperture photometry routines in DAOPHOT
(\citealt{stetson1987})  and then matched to form colors.  Currently, the
photometry for these three clusters is uncalibrated; however, the color
terms for the FORS2 array are small ($\sim$0.03 in $V-I$) and thus have
little effect on the shape of the MSTO/RC region, which spans a color
range of only $\sim$0.6 mag in $V-I$.  Regarding NGC 2193, the one cluster
in our initial sample with unpublished {\it HST} WFPC2 photometry, we
retrieved F450W and F555W images from the HST archive. These pipeline
processed images were photometered via the procedure outlined by
\citet{sarajedini1998}, including the \citet{holtzmanetal1995}
transformation coefficients.  Since the photometric zero points for WFPC2
are relatively uncertain, and the FORS2 data are uncalibrated, we proceed
with MSF as follows.  Utilizing the Z = 0.008 ([Fe/H] $\approx -$0.4) and
Z = 0.004 ([Fe/H] $\approx -$0.7) theoretical models from the Padova group
(\citealt{girardietal2002}), which include treatment for core overshoot,
we first shift the isochrones vertically to match the brightness of the RC
and then move them horizontally to match the color of the unevolved main
sequence.  For illustrative purposes, NGC 1651 and NGC 2173 are shown in
Fig.~\ref{fig:iso_fit}, with the Z = 0.008 isochrones over plotted for
log(Age)  = 9.25 and 9.30 for NGC 1651 and 9.15, 9.20, and 9.25 for NGC
2173;  based on these fits, we adopt ages of log(Age) = 9.28 (1.91 Gyr)
and 9.20 (1.58 Gyr) for NGC 1651 and NGC 2173, respectively, and we
estimate the error in our fits to be $\pm$ 0.05 in terms of log(Age).  
Table \ref{tab:cluster_ages} gives MSF ages for all clusters in our
preliminary sample with available optical photometry.  While neither NGC
1783 nor NGC 1978 has reliable photometry available in the literature,
both have ages determined by \citet{geisleretal1997}, who used the
difference in $V$-band magnitude between the cluster's RC and main
sequence turnoff to estimate cluster ages.  For clusters in common, we
find an offset of 0.03 in log(Age), where our MSF ages are younger than
their ages.  Therefore, for NGC 1783 and NGC 1978, we offset the values in
\citet{geisleretal1997} and adopt these as the ages for NGC 1783 and NGC
1978.

\section{Apparent and Absolute $K$-band RC Magnitudes}
\label{phot:sec:mags}

To calculate the apparent and absolute RC magnitudes, we generally follow
the method prescribed by GS02.  They determine the apparent $K$-band
magnitude of the RC ($\krc$) by placing a standard sized box (0.8 mag in
$K$ and 0.2 mag in $J-K$) around the RC; the median value of all stars
within this box is taken as $\krc$.  A constant box size is used in
conjunction with the median magnitude of the RC in an effort to eliminate
any selection effects that may occur in choosing the location of the box,
as well as to limit the effects of outliers on $\krc$.  In a few cases, we
have had to shift the box center slighty in color so as to avoid
contamination from RGB stars.  For predicting the absolute RC magnitude
($\mkrc$), GS02 combined available 2MASS photometry ($JK_S$)  for 14
Galactic open clusters, which also have internally consistent ages,
abundances, and distances, with an interpolation routine based on low
order polynomials.  The interpolation over the open clusters allows the
prediction of $\mkrc$ for a target cluster with a known age and [Fe/H].  
This method was applied to NGC 2158 by GS02 and to Hodge 4 and NGC 1651 by
\citet{sarajedinietal2002}, all with promising results.

Given ISPI's large field of view, before we can measure $\krc$ we must
separate the cluster stars from the field by performing radial cuts on our
data.  Where available, we use the cluster radii as determined by
\citet{grocholski2006}, which were based on the kinematics of individual
stars; typically, the farthest star from the cluster center that is moving
at the velocity of the cluster denotes the adopted radius.  For the four
clusters not in common with their study, radial cuts were chosen by eye,
using a combination of cluster images and our photometric catalogs.  We
note that small variations in the adopted cluster radii have no
appreciable effect on our results; a change in radius of $\pm$ 100 pixels
($\sim$ 0.5 arcmin) results in a change in $\mkrc$ of $\sim$ 0.03 mag.  
In Fig.~\ref{fig:rc_1}, we present the resulting $K$ vs.~$J-K$ cluster
CMDs, which extend from the tip of the RGB to $\sim$1.5 mag below the
helium burning RC; the standard size box used in calculating $\krc$ is
shown.  For each cluster, the measured value of $\krc$ is given in column
2 of Table \ref{tab:rcdist}, along with the standard error of the median
(column 3) and number of RC stars in each box (column 4).

Ideally, we would like to predict $\mkrc$ using the open cluster data
presented in GS02.  In practice, however, this is difficult since our LMC
cluster sample falls outside of the parameter space (in metallicity)
covered by the open clusters; tests of an extrapolation routine applied to
the target cluster abundances proved to be unreliable.  Instead, we turn
to the theoretical models of \citet[][see also
\citealt{girardietal2000}]{girardisalaris2001}, which provide expected
values of $\mkrc$ that span a large range of ages and metallicities and
encompass our LMC target clusters.  GS02 tested their open cluster data
against these theoretical models and found good agreement, with all
clusters lying within 1.5$\sigma$ of the appropriate model and no
systematic offset.  Since their comparison was based on data from the
Second Incremental Data Release of the 2MASS Point Source Catalog, we have
recompared the models and the data, using the updated 2MASS All Sky Data
Release.  With the new 2MASS photometry, we still find good agreement with
the models, however, there is now an offset of 0.08 mag, in that the
observed RC values are brighter than what is predicted by the models.  We
discuss this in more detail in \S \ref{phot:sec:errors}.  Given the ages
and metallicities listed in Table \ref{tab:cluster_ages}, we are able to
determine $\mkrc$ for each LMC cluster by interpolating over the
\citet{girardisalaris2001} models;  predicted values of $\mkrc$ are
presented in Table \ref{tab:rcdist}.  The quoted error in $\mkrc$ is
calculated by adding in quadrature the effects of age and abundance errors
on the predicted absolute RC magnitude.  We note that the five youngest
clusters in our sample have relatively large error bars due to the fact
that their ages place them in a region where the RC brightens rapidly with
increasing age (see Fig.~4 in GS02); thus, small errors in age result in
large errors in $\mkrc$.

\section{Cluster Distances and the Distance to the LMC}
\label{phot:sec:moduli}
\subsection{Absolute Distance Moduli}

With $\krc$ and $\mkrc$ in hand, cluster reddenings are all that is needed
to calculate absolute distance moduli.  The extinction maps of both
\citet{bursteinheiles1982} and \citet*{schlegeletal1998} cover the entire
LMC; however, \citet{schlegeletal1998} were not able to resolve the
temperature structure in the inner portions of the LMC and, therefore,
could not estimate the reddening reliably.  For most clusters, the two
reddeining maps give values in good agreement, although as some of our
clusters lie in the unresolved region, we adopt $\ebv$ values solely from
\citet{bursteinheiles1982} and assume an error of 20\%.  Reddenings are
converted to $A_K$ using the extinction law of \citet*{cardellietal1989},
where $R_V = 3.1$ and $A_K = 0.11 A_V$.  We note that, since $A_K$ is
approximately one third of $\ebv$, any differences between the two
extinction maps are ultimately negligible.  We also note that the adopted
values of $A_K$ are typically on the order of the error in measuring
$\krc$. In Table \ref{tab:rcdist} we give $\ebv$ and $A_K$ for the cluster
sample.  With absolute and apparent RC magnitudes and reddenings for each
cluster, absolute distance moduli, $\mmo$, are readily calculated and are
listed in Table \ref{tab:rcdist} along with the distance errors, which are
found by adding in quadrature the errors in $\krc$, $\mkrc$, and $\ebv$.


\subsection{LMC Cluster Distribution}
\label{phot:sec:distrib}

It has long been known that the disk of the LMC is inclined with respect
to the plane of the sky (see e.g., \citealt{caldwellcoulson1986}), and
this inclination is an important effect when using individual stars (or
clusters) to determine the distance to the LMC center.  Recent work using
field stars as a tracer of the disk (tip of the RGB and AGB,
\citealt{vdmcioni2001}; field RC stars, \citealt{olsensalyk2002}; carbon
stars, \citealt{vdmetal2002}; Cepheid variables,
\citealt{nikolaevetal2004}) has shown that the LMC has an inclination of
$i \sim31\degr-36\degr$, with a position angle of the line of nodes,
$\Theta$, between 120$\degr$ and 150$\degr$;  both of these quantities
have the standard definitions where $i = 0\degr$ for a face on disk and
$\Theta$ is measured counterclockwise from north.  The LMC centers adopted
by each of these authors, in addition to their derived values for $\Theta$
and $i$, are given in Table \ref{tab:geom}.  In Fig.~\ref{fig:schematic}
we plot the positions on the sky of our target clusters as well as the LMC
centers adopted by \citet[][ {\it filled square}]{vdmcioni2001}, \citet[][
{\it filled triangle}]{vdmetal2002}, and \citet[][ {\it filled
star}]{olsensalyk2002}. The solid lines passing through these points show
each author's position angle of the line of nodes.  We note that, for
clarity, we have not plotted the center and position angle of the line of
nodes from \citet{nikolaevetal2004} as they are very similar to the values
in \citet{olsensalyk2002}.  For reference, the 2$\degr$ near-infrared
isopleth (\citealt{vdm2001}), which roughly outlines the LMC bar, is
plotted as the dashed ellipse.  Conversion to Cartesian coordinates from
right ascension and declination was performed using a zenithal
equidistiant projection (e.g., \citealt{vdmcioni2001}, their
eqs.~[1]-[4]); lines of right ascension and declination have been marked
with dotted lines. In general, these geometries tell us that the northeast
portion of the LMC is closer to us than the southwest.  More specifically,
since points along the line of nodes are equidistant from the observer, in
the direction perpendicular to the line of nodes we would expect to see a
maximum gradient in cluster distance.

To compare our cluster distribution with the geometry of the LMC, in
Fig.~\ref{fig:dist_distrib} we plot cluster distance as a function of
radial distance along the line of maximum gradient. While we have used the
geometry of \citet{vdmcioni2001} to determine the position of the line of
maximum gradient, the choice in LMC geometry between these three recent
studies has little effect on the results (see \S \ref{phot:sec:lmcdist}).  
In the top panel, clusters are labeled for reference and in the bottom
panel we have included the 1$\sigma$ distance errors.  In addition, the
dashed line represents the disk of the LMC, where the LMC center ($x = 0$)
has a distance of 47.9 kpc (see \S \ref{phot:sec:lmcdist}) and $i =
34\fdg7$ (\citealt{vdmcioni2001}); the dotted line represents a constant
disk thickness of $\pm$ 1 kpc.  While a flared disk model
(\citealt{alvesnelson2000}) is probably a more correct representation of
the LMC's disk, for the purposes of our comparison a constant thickness
disk model is adequate.  Regardless, Fig.~\ref{fig:dist_distrib} shows
that, with the exception of the youngest clusters, which have inherently
uncertain distances, our results are consistent with the idea that the LMC
clusters lie in the same inclined, thick disk as defined by a variety of
field populations.

A disk-like cluster distribution is as expected, based on the kinematics
of the cluster system (\citealt{schommeretal1992}), but this is the first
time it has been demonstrated that the clusters and field stars reside in
the same disk.  This result is in contrast to the recent findings of
\citet{kerberetal2006}, who used the MSTO to calculate distances for 15
LMC clusters.  From their data they found a disk-like distribution for
their clusters, along with an inclination of $39\degr \pm 7\degr$, which
is $\sim 8\degr$ steeper than the $30\fdg7\pm1\fdg1$ disk inclination that
\citet{kerberetal2006} adopted from \citet{nikolaevetal2004}.  
\citet{kerberetal2006} interpreted this inclination difference as
suggesting that the LMC's intermediate-age clusters formed in a different
disk than the field stars.  However, they discuss neither the results of
\citet{vdmcioni2001} nor \citet{olsensalyk2002}, who find disk
inclinations of $34\fdg7\pm6\fdg2$ and $35\fdg8\pm2\fdg4$, respectively,
both in agreement with the cluster disk inclination found by
\citet{kerberetal2006}.

We note in passing that \citet{olsensalyk2002} found what appears to be a
warp in the southwest portion of the LMC.  Their fields in this region are
brighter than expected, giving the impression that they have been pulled
toward the MW.  There is, however, a possible problem with the reddening
corrections that \citet{olsensalyk2002} have applied to these fields, which
may explain the apparent warp.  As only two of our target clusters, NGC
1651 and SL 61, lie in the warped area, we are not in a position to
comment on their result.

Since galactic disks are relatively fragile, and it is highly unlikely
that clusters would form in a halo and then be perturbed into a disk, the
disk-like distribution and kinematics of our LMC clusters suggest that
they formed in a disk.  As ESO 121 is the oldest cluster in our IR sample,
its residence in the LMC's disk implies that the disk formed $\sim$9 Gyr
ago.  However, ESO 121 is well known to be the only cluster in the LMC
with an age between approximately 3 Gyr and 13 Gyr.  To further explore
the age of the disk, we turn to the LMC's {\it bona fide} old ($\sim$13
Gyr) globular cluster population and the optical photometry of
A.~R.~Walker (see \citealt{walker1985,walkermack1988,walker1989,
walker1990, walker1992_retic,walker1992_1466,walker1993}).  Walker
measured the mean apparent $V$-band magnitude ($V_{RR}$) of RR Lyrae stars
in seven LMC globular clusters and, using their pulsational properties,
was able to estimate cluster metallicities.  Given the metallicity of a
cluster, the mean absolute RR Lyrae magnitude is determined by $M_{V}^{RR}
= 0.23 [Fe/H] + c$ (\citealt{chaboyer1999}), and by adopting reddenings
from \citet{bursteinheiles1982} we can readily calculate distances for
these seven clusters.  The zeropoint, $c$, in the above relation is chosen
such that NGC 1835 lies on the dashed line.  Cluster information is given
in Table \ref{tab:old_clust}, and these new data points are plotted in
Fig.~\ref{fig:dist_distrib} as open circles, along with their 1$\sigma$
errors.  The errors in [Fe/H] and $V_{RR}$ are taken from Walker and we
assume a 20\% error in $\ebv$ for all clusters except Reticulum, for which
we adopt 0.02 mag.  Fig.~\ref{fig:dist_distrib} shows that, like the
intermediate age clusters, the old globular clusters are distributed in a
manner that is consistent with the thick, inclined disk geometry of the
LMC field stars.  The agreement between the old globular clusters and the
disk suggest that cluster like NGC 2257 and NGC 1466 formed in, and still
reside in, the disk.  The disk of the LMC, therefore, must be roughly the
same age as the globular clusters, $\sim$13 Gyr old.

Lastly, we note the position of NGC 1841.  This cluster resides $\sim$12
kpc from the LMC center (to the south), which places it near the tidal
radius ($r_t = 15.0 \pm 4.5$ kpc, \citealt{vdmetal2002}) of the LMC, and,
as can be seen in Fig.~\ref{fig:dist_distrib}, it sits well out of the
plane of the disk, in the direction of the Milky Way.  Thus, NGC 1841 is
likely to have either been pulled out of the disk, or stripped from the
LMC altogether, in a close encounter with the Milky Way.

\subsection{The Distance to the LMC Center}
\label{phot:sec:lmcdist}

For any given point, $P$, that resides in the disk of the LMC, the 
distance, $D$, of 
that point is related to the distance to the center of the LMC, 
$D_0$, by
\begin{equation}
D/D_0 = \cos i/[\cos i \cos \rho - \sin i \sin \rho \sin (\phi - \theta)],
\label{eq:phot:distratio}
\end{equation}
where $i$ is the inclination of the disk and $\theta = \Theta + 90$ (see 
\citealt{vdmcioni2001} for a detailed discussion of equations 
\ref{eq:phot:distratio}$-$\ref{eq:phot:phitwo}).  
The angular coordinate $\rho$ is defined as the  
angular separation on the sky between $P$ and the LMC 
center, while 
$\phi$ is the position angle of $P$ relative to the center.  
Typically, $\phi$ is measured
counterclockwise from the axis that runs in the direction of decreasing 
right ascension and passes through the LMC center.  
These coordinates ($\rho$, $\phi$) can be uniquely defined by the cosine 
and sine rule of spherical trigonometry and the analog formula, which give  
\begin{equation}
\cos \rho = \cos \delta \cos \delta_0 \cos (\alpha - \alpha_0) + \sin
\delta \sin \delta_0, 
\label{eq:phot:rho}
\end{equation}
\begin{equation}
\sin \rho \cos \phi = - \cos \delta \sin (\alpha - \alpha_0),
\label{eq:phot:phione}
\end{equation}
and
\begin{equation}
\sin \rho \sin \phi = \sin \delta \cos \delta_0 - \cos \delta \sin
\delta_0 \cos (\alpha - \alpha_0).
\label{eq:phot:phitwo}
\end{equation}  
In equations \ref{eq:phot:rho}$-$\ref{eq:phot:phitwo}, $\alpha_0$ and 
$\delta_0$ are the right ascension and declination of the LMC center while 
$\alpha$ and $\delta$ mark the position on the sky of $P$.  Therefore, 
since 
it is reasonable to assume that our target clusters lie in the disk of the 
LMC, as defined by the field stars (\S \ref{phot:sec:distrib}), we can use 
the distances of our clusters in conjunction with the LMC geometry to 
calculate the distance to the center of the LMC.

As an example, we adopt $i = 34\fdg7$ and $\Theta = 122\fdg5$
(\citealt{vdmcioni2001}), and calculate values for the LMC center distance
based on the distance and position of each of our 17 target clusters.  
Raw cluster distances from Table \ref{tab:rcdist} and the corresponding
LMC distance are given in Table \ref{tab:cen_dist} with the LMC distance
errors calculated by propogating the errors in $i$, $\Theta$, and $D$
through equation \ref{eq:phot:distratio}.  Finally, we calculate the
distance to the LMC as the mean of the individual center distances, for
which we find $D_0 = 47.9 \pm 0.9$ kpc, or $\mmo = 18.40 \pm 0.04$; the
quoted error is the standard error of the mean.  We note that, while
calculating the straight mean does include the young clusters, which have
uncertain distances, we have found that the mean, median, weighted mean,
and 2$\sigma$ clipped mean all give distances within 0.01 mag of each
other, thus we have chosen to simply adopt the mean as our final distance.  
In addition to \citet{vdmcioni2001}, we also use the geometry of
\citet{olsensalyk2002}, \citet{vdmetal2002}, and \citet{nikolaevetal2004}
to calculate the distance to the LMC, with all four mean distances given
in Table \ref{tab:geom}. The final distances, $D_0 =$ $47.9 \pm 0.9$ kpc,
$48.1 \pm 0.9$ kpc, $47.9 \pm 0.9$ kpc, and $48.1 \pm 0.9$ kpc, are all in
excellent agreement, which shows that the choice of geometry between these
four authors has little effect on the distance to the LMC center.

\subsection{Systematic Errors}
\label{phot:sec:errors}

An analysis of our approach to calculating cluster distances gives two
possible sources of systematic errors.  The first source of error in our
calculations arises from our interpolation method.  As discussed in \S
\ref{phot:sec:mags}, due to the location of our target clusters in the
age-metallicity parameter space, we are not able to interpolate over the
open cluster data in GS02.  Instead, we have had to use the
theoretical models of \citet{girardisalaris2001} for our interpolation.  
While the models are in good agreement with the open cluster data, they
predict absolute magnitudes that are, on average, 0.08 mag fainter than
what is observed.  An additional systematic error may arise from our
choice of reddening map.  \citet{bursteinheiles1982} zeropoint their
reddening maps to an area near the north galactic pole which was long
believed to be a direction of zero reddening.  \citet{schlegeletal1998},
however, find $\ebv = 0.02$ mag for the same location on the sky.  These
two systematic errors work in opposite directions; if we applied a
correction for the interpolation error, clusters would move {\it closer},
while a correction for the reddening error would make them appear farther
away.  However, since $A_K = 0.341\ebv$, the systematic reddening error is
small and is dominated by the systematic error due to our interpolation.  
Therefore, we adopt 0.08 mag as our systematic error.

\section{Comparison to Previous Distances}
\label{phot:sec:dist}

Since an extensive review of LMC distances determined by a variety of
standard candles can be found in \citet{clementinietal2003}, herein we
restrict our comparison to only a couple recent distance calculations.  
The only previous LMC cluster distances based on the $K$-band luminosity
of the RC are presented in \citet{sarajedinietal2002} and, using the
approach described in GS02, they find $\mmo = 18.55 \pm 0.12$ and $18.52
\pm 0.17$ for NGC 1651 and Hodge 4, respectively.  Both distances are
farther than what we find for the same clusters, due primarily to their
photometric calibration.  For both clusters, \citet{sarajedinietal2002}
measure $\krc$ to be $\sim$0.1 mag fainter than our values.  Given the
small number of standard stars used by \citet{sarajedinietal2002} along
with their small field of view, which provided only a handful of stars for
aperture correction determination, this difference in photometric
zeropoint is not unexpected.

Most recently, \citet{macrietal2006} observed Cepheid variables in two
fields in the maser-host galaxy NGC 4258.  By comparing the LMC's Cepheid
P-L relation to their observations of variables in NGC 4258,
\citet{macrietal2006} were able to calculate a {\it relative} distance
between these two galaxies of $\Delta \mmo = 10.88 \pm 0.04$ (random) $\pm
0.05$ (systematic).  Being a maser-host galaxy, NGC 4258 has an accurate
geometric distance ($29.29 \pm 0.09 \pm 0.12$ mag) that, combined with the
Cepheid-based relative distance, allowed \citet{macrietal2006} to
calculate the distance to the LMC.  They find $\mmo = 18.41 \pm 0.10 \pm
0.13$, in excellent agreement with our results.  As discussed by Marci et
al.~(2006), this improved distance has implications for calculations of
$H_0$.  The {\it HST} Key Project to determine the Hubble constant (see
\citealt{freedmanetal2001}) adopted $\mmo = 18.5 \pm 0.1$ as their
distance to the LMC, which acts as the zeropoint for the extragalactic
distance scale.  Using this longer distance, \citet{freedmanetal2001} find
$H_0 = 72 \pm 8$ km s$^{-1}$ Mpc$^{-1}$.  In recalculating $H_0$,
\citet{macrietal2006} find that the shorter LMC distance increases the
Hubble constant $\sim$3\%.  However, they find that their new coefficient
of metallicity dependence for Cepheid variables has the opposite effect,
changing $H_0$ by $\sim-$2\%.  Thus, the cumulative effect results in only
a small change in the Hubble constant.  With their new results, they
calculate $H_0 = 74 \pm 3 \pm 6$ km s$^{-1}$ Mpc$^{-1}$.

\section{Summary}
\label{phot:sec:summary}

In this paper we have presented results of a near-infrared photometric
study of populous clusters in the LMC.  Using ISPI on the CTIO 4m we
obtained $JK'$ photometry down to $\sim$1.5 mag below the core helium
burning red clump stars in 17 clusters, allowing us to accurately measure
the {\it apparent} $K$-band magnitude of the RC.  In a similar approach to
that of GS02, we combine cluster ages and metallicities with
theoretical models to predict the {\it absolute} $K$-band RC magnitude for
each of these clusters.  Thus, we are able to determine accurate cluster
distances and explore the 3-dimensional cluster distribution as well as 
calculate the distance to the center of the LMC.  The main results of our 
paper are as follows:

1) We have compiled deep optical photometry (below the MSTO) for 15 of our
clusters.  By combining these data with previously published
metallicities, we are able to break the well known age-metallicity
degeneracy and calculate accurate cluster ages via MSTO fitting with
theoretical isochrones that include treatment for core overshoot.  The
intermediate age clusters range in age from only $\sim$1-3 Gyr; thus,
these MSF ages do not close the LMC's cluster age gap.  We confirm that
ESO 121, the only LMC cluster known to have an age between $\sim$3-13 Gyr,
formed approximately 9 Gyr ago.

2) By combining $\krc$ measured from our near IR photometry with the
values of $\mkrc$ predicted by theoretical models, we have determined
accurate distances for all 17 clusters in our sample; our average standard
error of the mean distance is 0.08 mag, or 1.8 kpc.  This work represents
the largest sample of LMC clusters with distances derived in an internally
consistent way. 


3) The cluster distances allow us to explore the spatial
distribution of the LMC cluster system.  Previous work has shown that the
LMC field populations lie in a thick, inclined disk and our results
illustrate that the clusters are distributed in the same manner.  A
disk-like distribution for all LMC clusters has been inferred from the
kinematics of the cluster system, however, our results mark the first time
that it has been demonstrated that the clusters and the field stars lie in 
the same plane.


4) Previously published RR Lyrae data for seven old globular clusters have
allowed us to calculate distances for these clusters and compare their
distribution to the geometry of the LMC.  Like the intermediate age
clusters, the globular clusters have a distribution that is consistent
with residence in the disk of the LMC.  

5) Given that it is unlikely for clusters to form in a halo and then be
perturbed {\it into} a disk, the disk-like kinematics and distribution of
the LMC clusters implies that they formed in a disk.  The fact that old
clusters (e.g.,~NGC 2257, NGC 1466, ESO 121) are seen to currently reside
in the disk suggests that they also formed in the LMC's disk.  From this,
we infer that the disk of the LMC must have formed about $\sim$13 Gyr ago.


6) The old globular cluster NGC 1841 lies near the LMC's tidal radius
and well out of the plane of the disk, in the direction of the Milky Way.
Its position suggests that it was pulled out of the disk, or possibly
stripped from the LMC, in a close encounter with the Milky Way. 

7) Taking the inclined geometry of the LMC into account, we find the mean
distance to the center of this nearby galaxy to be $\mmo = 18.40 \pm 0.04
\pm 0.08$ or $D_0 = 47.9 \pm 0.9 \pm 1.8$ kpc.  Our result is in excellent
agreement with the recent work of \citet{macrietal2006} who found $\mmo =
18.41 \pm 0.1 \pm 0.13$ by comparing Cepheid variables in the maser-host
galaxy NGC 4258 with those in the LMC.  This distance, however, is
$\sim$0.1 mag shorter than the commonly accepted distance of $18.5 \pm
0.1$ mag, which was used in the {\it HST} Key Project to calculate $H_0$
(see \citealt{freedmanetal2001}).  This shorter distance has the 
effect of increasing $H_0$ by $\sim$3\% \citep{macrietal2006}.


\acknowledgments

This research was supported by NSF CAREER grant AST-0094048 to AS.  We
would like to thank Mike Barker for assistance with the near-IR
observations, Andy Stephens for providing us with a copy of his data, and 
Steve Eikenberry for helpful discussions regarding the data processing.  
The authors appreciate the helpful comments of an anonymous referee.

\clearpage

\begin{figure}[!t]
\begin{center}
\plotone{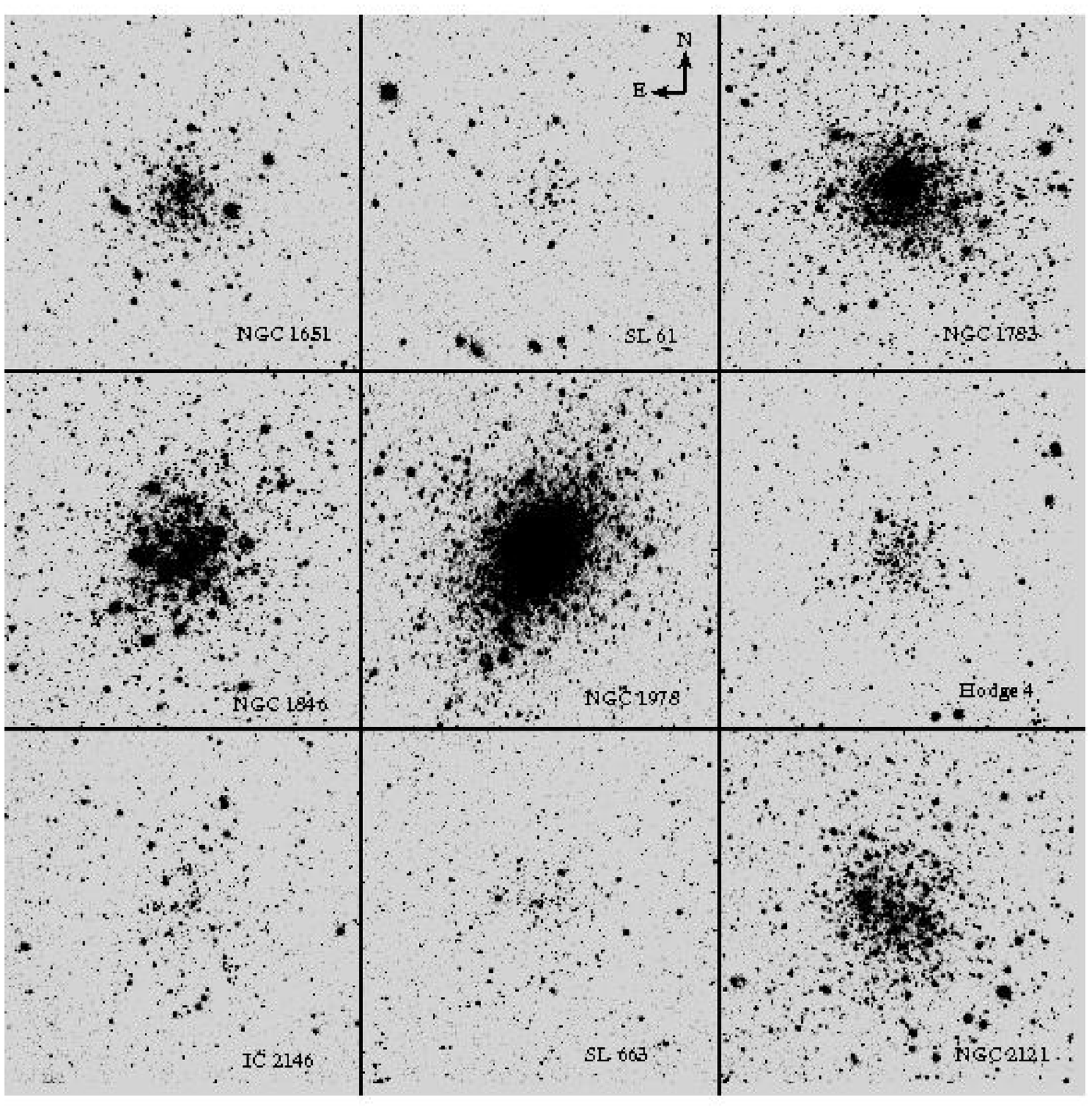}
\caption{$K'$-band images for 
all target clusters.  We have used the 
final combined long exposures and selected a region $\sim 4\arcmin \times 
4\arcmin$ in size around each cluster.  In all frames, clusters are 
labeled and the orientation is such that north is up and east is to the 
left.
}\label{fig:kbandphot}
\end{center}
\end{figure}
\addtocounter{figure}{-1}
\begin{figure}[!t]
  \begin{center}
\plotone{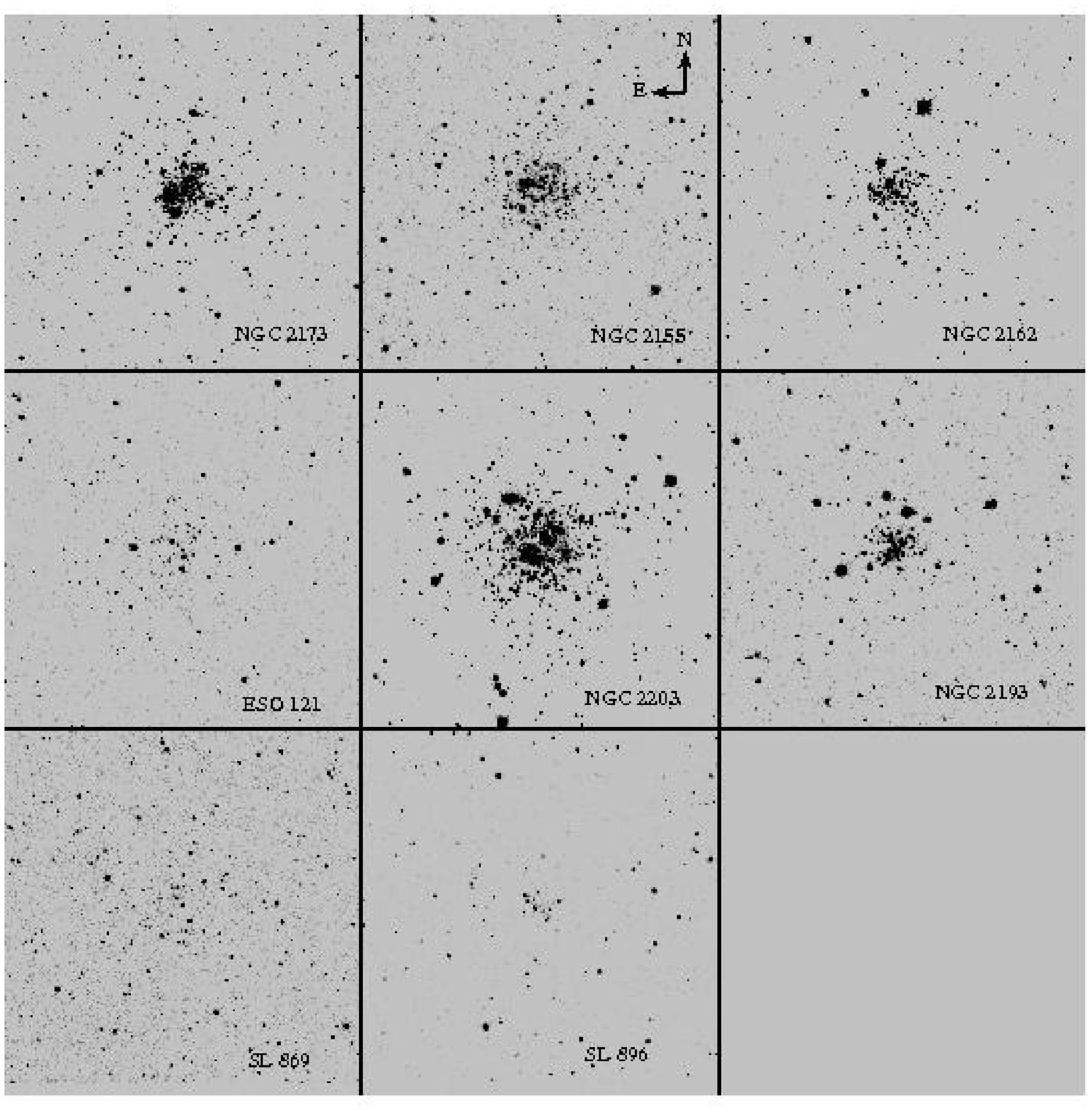}
\caption{- Continued.
}\label{fig:kbandphot2}
\end{center}
\end{figure}

\clearpage

\begin{figure}
\begin{center}
\plotone{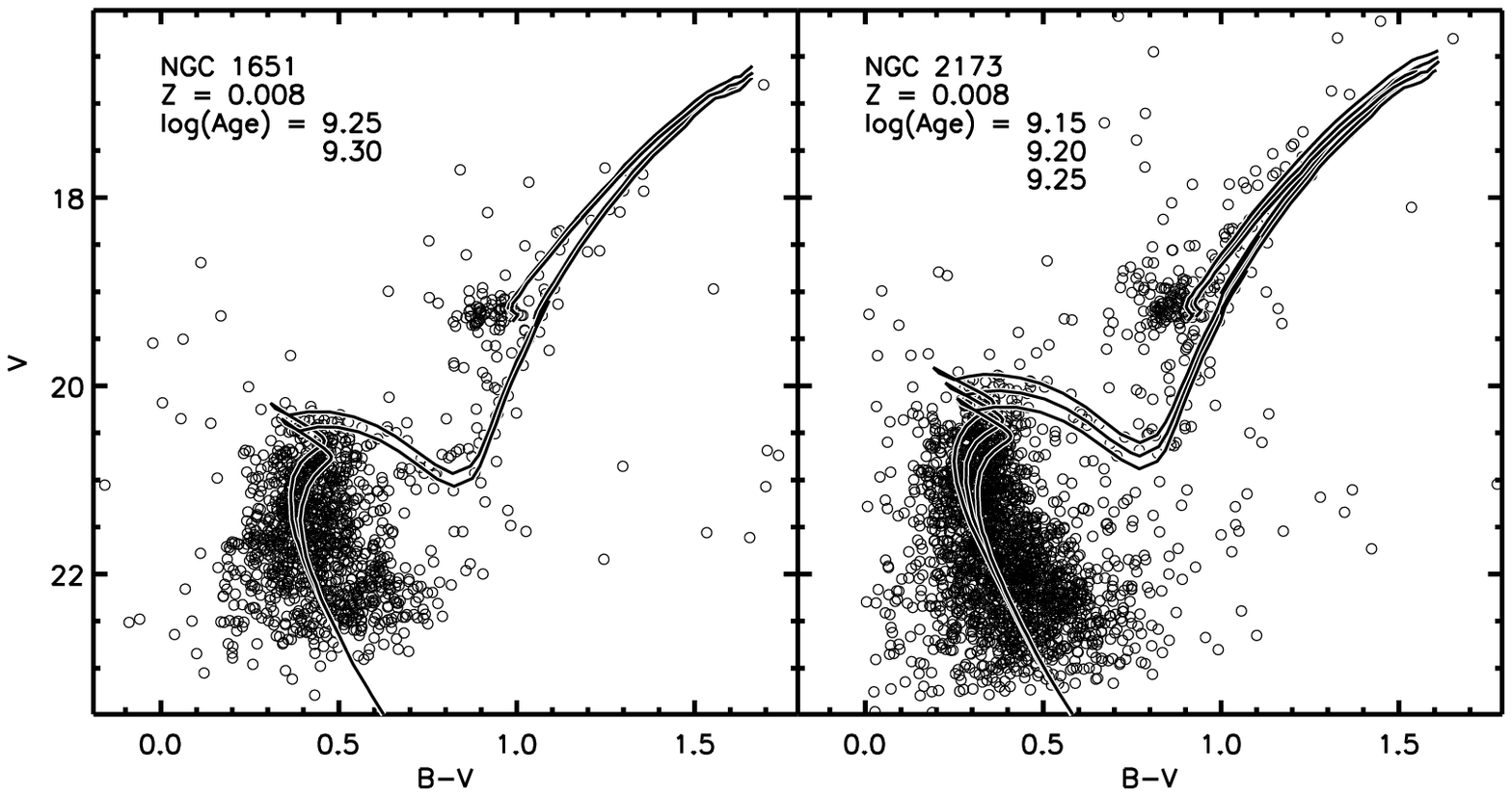}
\caption{Optical photometry for NGC 1651 (left) and NGC 2173 (right), 
overplotted with the Z = 0.008 theoretical isochrones from 
\citet{girardietal2002}; isochrone ages are listed in the figure.  These 
plots illustrate our MSF method where we match isochrones to the 
brightness of 
the RC and color of the unevolved main sequence to determine cluster ages.  
}\label{fig:iso_fit}
\end{center}
\end{figure}

\clearpage

\begin{figure}
\begin{center}
\epsscale{.90}
\plotone{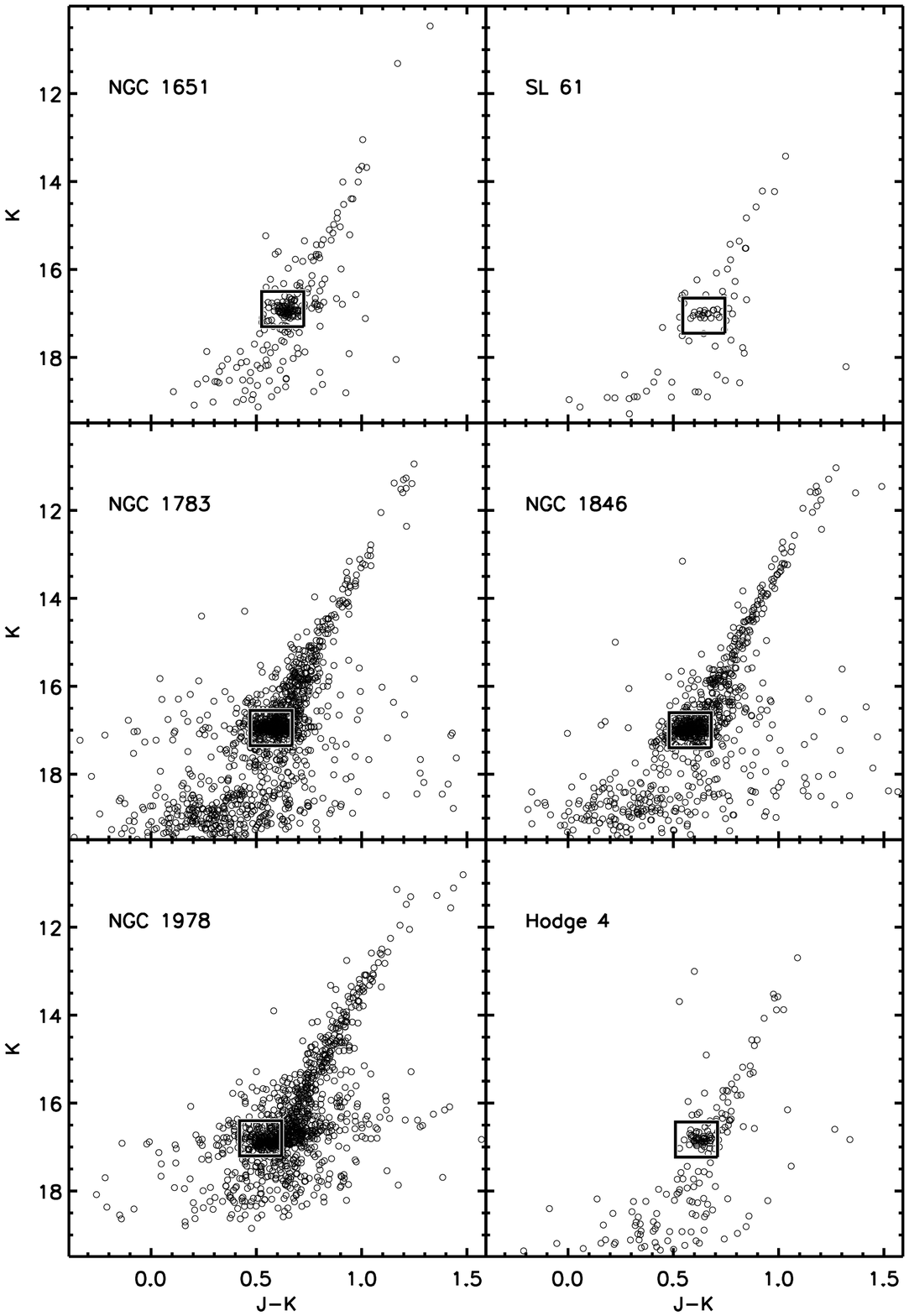}
\caption{Near-infrared CMDs for the 17 clusters in our sample.  Cluster 
RCs are denoted by the box and all stars within this box are used in 
calculating $\mkrc$. 
}\label{fig:rc_1}
\end{center}
\end{figure}
\addtocounter{figure}{-1}
\clearpage

\begin{figure}
\begin{center}
\epsscale{.90}
\plotone{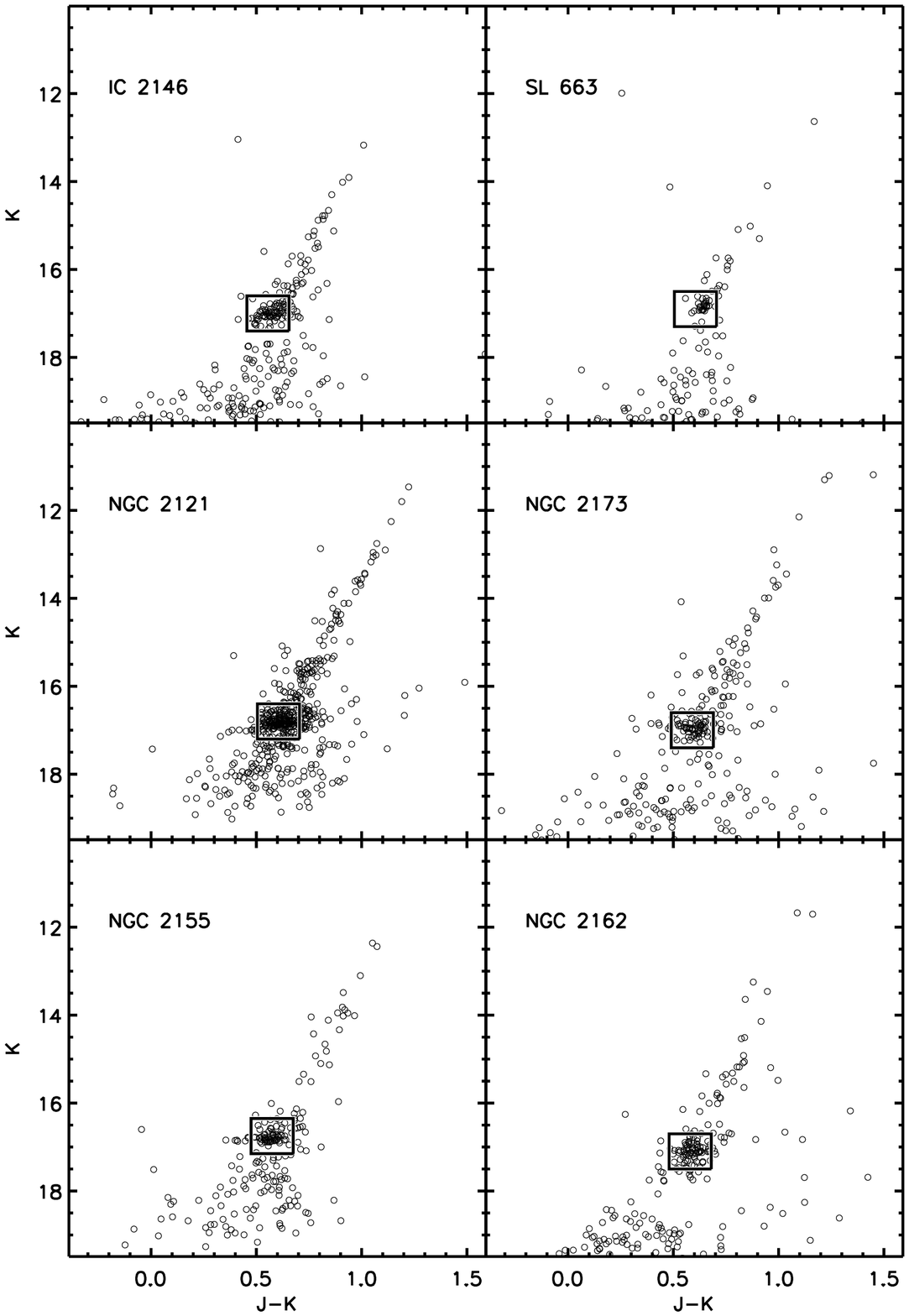}
\caption{{\it Continued.}
}\label{fig:rc_2}  
\end{center}
\end{figure}
\addtocounter{figure}{-1}
\clearpage

\begin{figure}
\begin{center}
\epsscale{.90}
\plotone{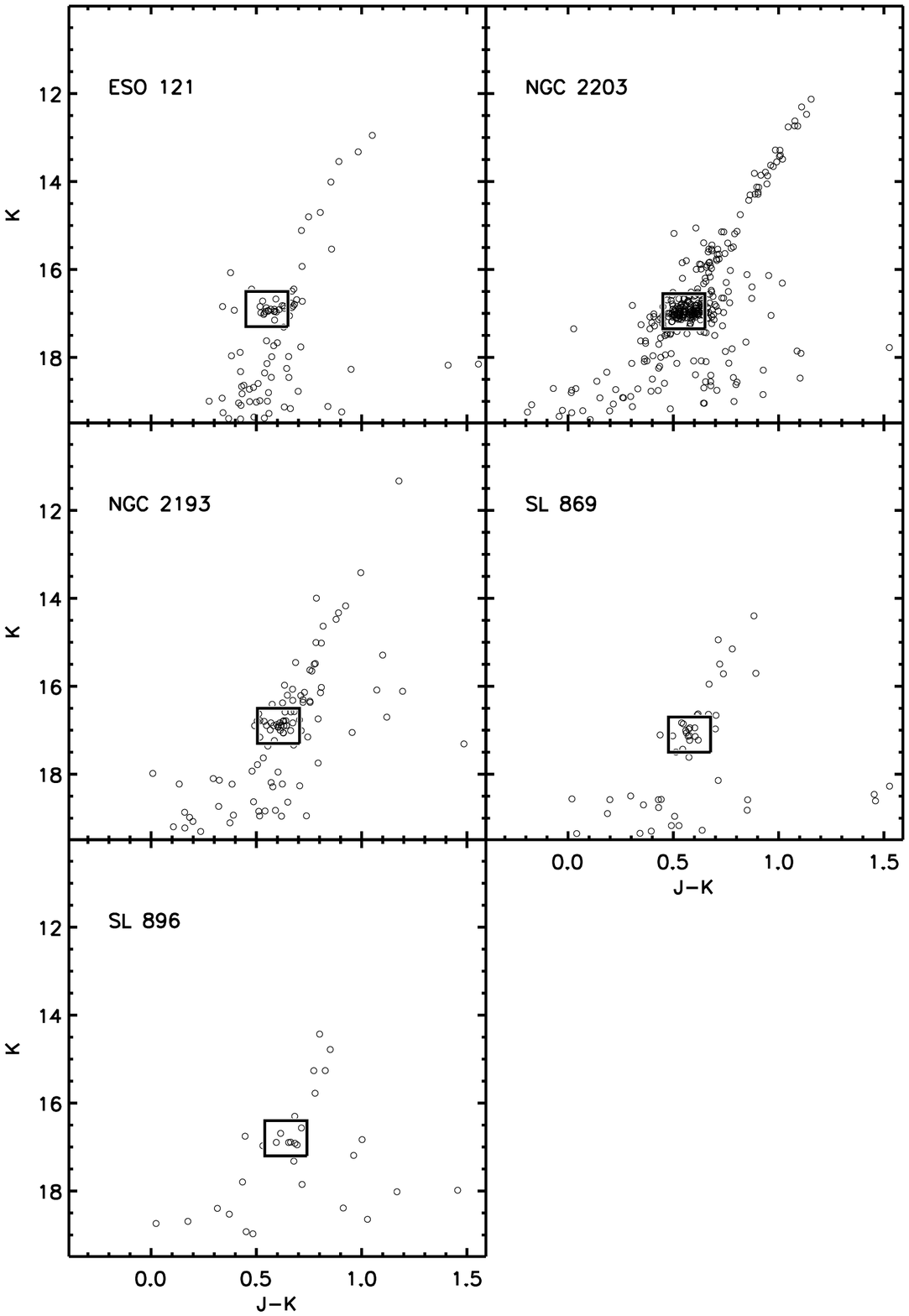}
\caption{{\it Continued.}
}\label{fig:rc_3}  
\end{center}
\end{figure}

\clearpage

\begin{figure}  
\begin{center}
\epsscale{1.0}
\plotone{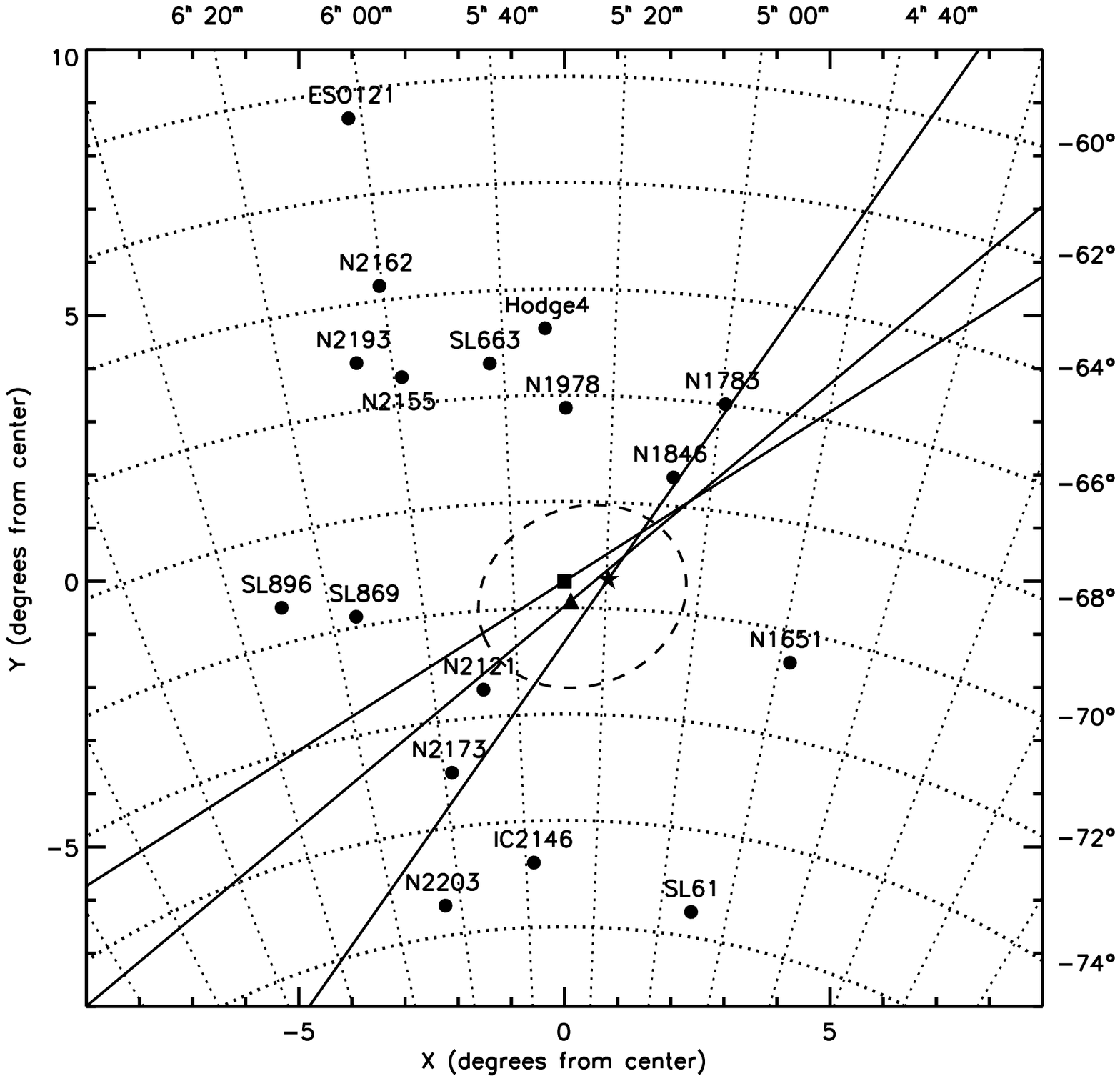}
\caption{Schematic diagram showing the positions on the sky of our target 
clusters.  The dashed ellipse represents the 2$\degr$ near-infrared 
isopleth from \citet{vdm2001}, which roughly outlines the LMC's bar.  
Also shown are the LMC centers used by 
\citet[][ {\it filled square}]{vdmcioni2001}, 
\citet[][ {\it filled triangle}]{vdmetal2002},
and \citet[][ {\it filled star}]{olsensalyk2002}.
The position 
angle of the line of nodes derived by each of these authors is plotted as 
the solid line passing through the appropriate LMC center.
 }\label{fig:schematic}
\end{center}
\end{figure}

\clearpage

\begin{figure}
\begin{center}
\epsscale{.65}
\plotone{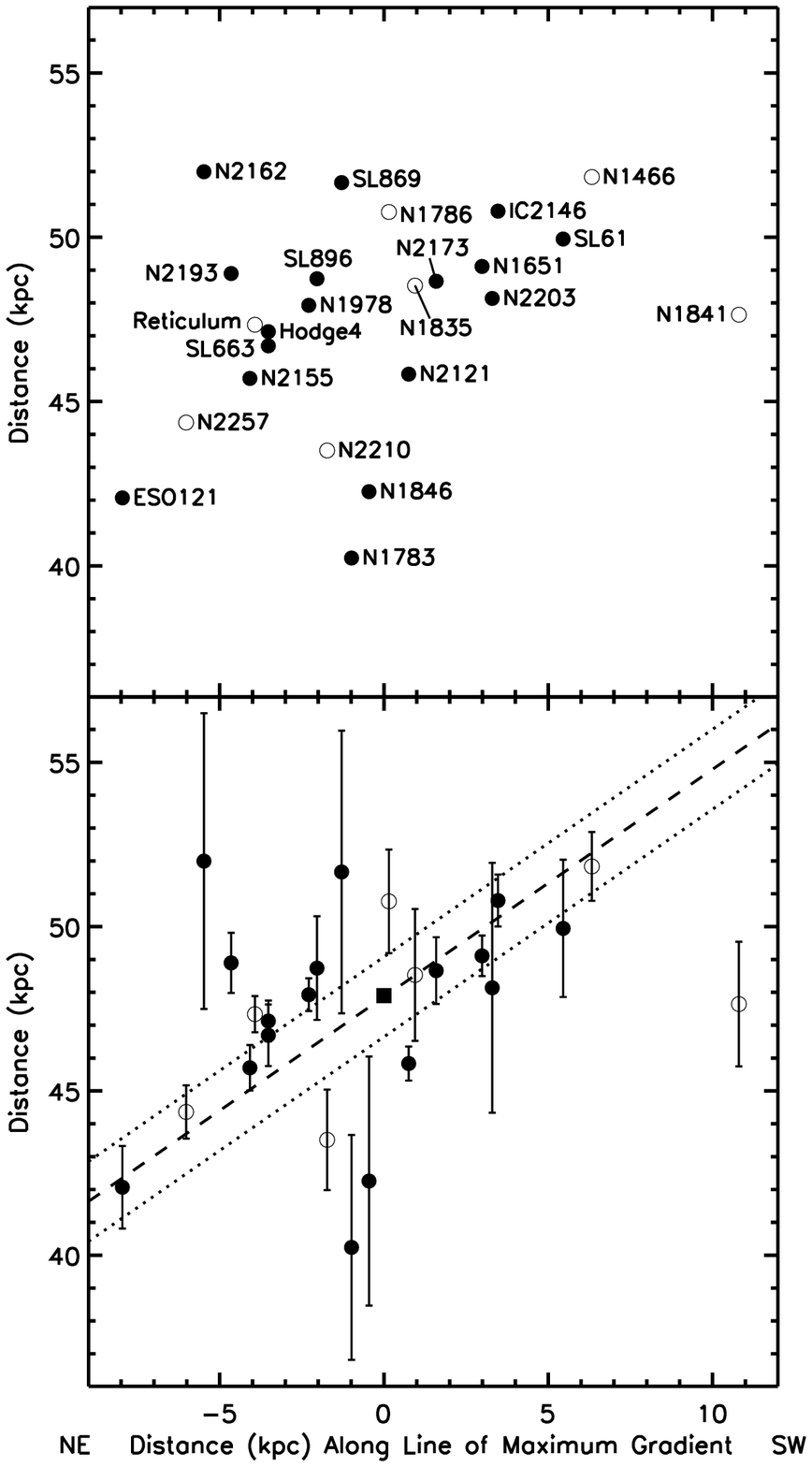}
\caption{Cluster distances as a function of their 
position along the line of maximum gradient (see \S 
\ref{phot:sec:distrib}).  Open circles mark the old globular clusters from 
Walker while the filled circles represent the populous clusters in our 
study.  In the bottom panel, the dashed line marks the LMC's disk with $i 
= 34\fdg7$ and $D_0 = 47.9$ kpc (at $x$ = 0), and the dotted lines 
represent a disk thickness of $\pm$ 1 kpc; the filled square denotes the 
center of the LMC.  This plot illustrates that both the old and 
intermediate age clusters are distributed along the disk of the LMC.  
}\label{fig:dist_distrib}
\end{center}
\end{figure}

\clearpage

\begin{deluxetable}{lccccccccccc}
\tabletypesize{\scriptsize}
\tablecaption{Exposure Times at Each Dither 
Point\label{tab:exposure_times}}
\tablewidth{0pt}
\tablehead{
\colhead{Dates} & \colhead{J} &
\colhead{H} &
\colhead{K'}
}
\startdata
20-22 Jan 2003& 60s & 15s $\times$ 6 & 10s $\times$ 9 \\
06-08 Feb 2004& 4s, 20s, 36s & 4s, 15s $\times$ 6 & 4s, 10s $\times$ 9 \\
\enddata
\end{deluxetable}

\begin{deluxetable}{lccccccccccc}
\tabletypesize{\scriptsize}
\tablecaption{LMC Cluster Sample Information\label{tab:cluster_info}}
\tablewidth{0pt}
\tablehead{
\colhead{Cluster} & \colhead{Alternate} &
\colhead{R.A.} & \colhead{Decl.} & \colhead{Filters} & \colhead{Run}\\
& Name & (J2000.0) & (J2000.0)& &
}
\startdata

NGC 1651  & SL 7,  LW 12    &4 37 33& $-$70 35 08& $JHK'$ & 1,2\\ 
SL 61     & LW 79           &4 50 45& $-$75 32 00& $J...K'$ &2\\
NGC 1783  & SL 148          &4 59 09& $-$65 59 14& $J...K'$ &2\\
NGC 1846  & SL 243          &5 07 35& $-$67 27 31& $J...K'$ &2\\
NGC 1978  & SL 501          &5 28 45& $-$66 14 09& $JHK'$ &1,2\\
Hodge 4   & SL 556, LW 237  &5 32 25& $-$64 44 12& $JHK'$ & 1,2\\
IC 2146   & SL 632, LW 258  &5 37 46& $-$74 47 00& $J...K'$ &2\\
SL 663    & LW 273          &5 42 29& $-$65 21 48& $J...K'$ &2\\
NGC 2121  & SL 725, LW 303  &5 48 12& $-$71 28 52& $JHK'$ &1,2\\
NGC 2173  & SL 807, LW 348  &5 57 58& $-$72 58 41& $J...K'$ &2\\
NGC 2155  & SL 803, LW 347  &5 58 33& $-$65 28 35& $JHK'$ &1,2\\
NGC 2162  & SL 814, LW 351  &6 00 30& $-$63 43 19& $J...K'$ &2\\
ESO 121-03 &                &6 02 03& $-$60 31 26& $JHK'$ & 1,2\\
NGC 2203  & SL 836, LW 380  &6 04 43& $-$75 26 18& $J...K'$ &2\\
NGC 2193  & SL 839, LW 387  &6 06 18& $-$65 05 57& $JHK'$ &1,2\\
SL 869    & LW 441          &6 14 41& $-$69 48 07& $JHK'$ &2\\
SL 896    & LW 480          &6 29 58& $-$69 20 00& $JHK'$ &1,2
\enddata
\tablecomments{Units of right ascension are in hours, minutes, and seconds 
and units of declination are in degrees, arcminutes, and arcseconds.  
}

\end{deluxetable}

\clearpage

\begin{deluxetable}{lccccccccccc}
\tabletypesize{\scriptsize}
\tablecaption{LMC Cluster Ages and Metallicities\label{tab:cluster_ages}}
\tablewidth{0pt}
\tablehead{
\colhead{Cluster} & 
\colhead{[Fe/H]\tablenotemark{a}} &  
\colhead{$\sigma_{[Fe/H]}$\tablenotemark{a}} &
\colhead{Log Age} & \colhead{Age (Gyr)} & \colhead{CMD Ref.}
}
\startdata
NGC 1783   & $-0.47$\tablenotemark{b} & 0.14\tablenotemark{b} &
9.08\tablenotemark{d} & 1.20  & $-$\\
NGC 1846   & $-0.49$ & 0.03 & 9.10 & 1.26  & 7\\
NGC 2162   & $-0.46$ & 0.07 & 9.15 & 1.41  & 1\\
NGC 2203   & $-0.41$ & 0.03 & 9.15 & 1.41  & 7\\
SL 869     & $-0.40$ & 0.04 & 9.15 & 1.41  & 6\\
SL 61      & $-0.35$ & 0.04 & 9.18 & 1.51  & 4\\
NGC 2173   & $-0.42$ & 0.03 & 9.20 & 1.58  & 1\\
IC 2146    & $-0.41$ & 0.02 & 9.25 & 1.78  & 7\\
NGC 1978   & $-0.38$\tablenotemark{c} & 0.07\tablenotemark{c} &
9.27\tablenotemark{d} & 1.86  & $-$\\
NGC 1651   & $-0.53$ & 0.03 & 9.28 & 1.91  & 1\\
NGC 2193   & $-0.49$ & 0.05 & 9.30 & 2.00  & 3\\
Hodge 4    & $-0.55$ & 0.06 & 9.33 & 2.14  & 5\\
SL 896     & $-0.48$\tablenotemark{e} & 0.09\tablenotemark{e} & 9.33 & 
2.14  & 6\\
NGC 2155   & $-0.50$ & 0.05 & 9.45 & 2.82  & 1\\
SL 663     & $-0.54$ & 0.05 & 9.45 & 2.82  & 1\\
NGC 2121   & $-0.50$ & 0.03 & 9.48 & 3.02  & 5\\
ESO 121-03 & $-0.91$\tablenotemark{f} & 0.16\tablenotemark{f} & 9.95 & 
8.91  & 2
\enddata


\tablecomments{Optical photometry used to construct the CMDs comes from
the following sources: 
(1) \citet{brocatoetal2001}; 
(2) \citet{bicaetal1998};  
(3) HST GO-5475; 
(4) \citet{mateohodge1985}; 
(5) \citet{sarajedini1998}; 
(6) \citet{piattietal2002}; 
(7) Grocholski et al.~(2007, in prep) 
}

\tablenotetext{a}{From Grocholski et al.~(2006), unless noted.}
\tablenotetext{b}{From Cole et al.~(in prep)}
\tablenotetext{c}{From \citet{ferraroetal2006}}
\tablenotetext{d}{Ages adjusted from \citet{geisleretal1997}}
\tablenotetext{e}{Mean value of the intermediate metallicity clusters 
from Grocholski et al.~(2006)} 
\tablenotetext{f}{From \citet{hilletal2000}}
\end{deluxetable}

\begin{deluxetable}{lccccccccccc}
\tabletypesize{\scriptsize}
\tablecaption{Calculated Red Clump Values and Cluster 
Distances\label{tab:rcdist}}
\tablewidth{0pt}
\tablehead{
\colhead{Cluster} & \colhead{$\krc$} & 
\colhead{$\sigma_{\overline{\krc}}$} & \colhead{$n$} &
\colhead{$\mkrc$} & 
\colhead{$\sigma_{\mkrc}$} & \colhead{$\ebv$} & 
\colhead{$\ak$} & \colhead{$\mmo$} & \colhead{$\sigma_{\mmo}$} &
\colhead{D} & \colhead{$\sigma_{D}$}
\\
\colhead{Name} & & & Stars & & & & & & & (kpc) & (kpc) 
}
\startdata
NGC 1651 & 16.93 & 0.02 & 93  & $-$1.56 & 0.02 & 0.10 & 0.034 & 18.46 & 
0.03 & 49.1 & 0.6\\ 
SL 61    & 17.01 & 0.03 & 22  & $-$1.52 & 0.08 & 0.11 & 0.038 & 18.49 & 
0.09 & 49.9 & 2.1\\
NGC 1783 & 16.93 & 0.01 & 384 & $-$1.10 & 0.18 & 0.02 & 0.007 & 18.02 & 
0.18 & 40.2 & 3.4\\
NGC 1846 & 16.98 & 0.01 & 301 & $-$1.17 & 0.19 & 0.06 & 0.020 & 18.13 & 
0.19 & 42.3 & 3.8\\
NGC 1978 & 16.86 & 0.01 & 231 & $-$1.56 & 0.02 & 0.05 & 0.017 & 18.40 & 
0.02 & 47.9 & 0.5\\
Hodge 4  & 16.81 & 0.02 & 48  & $-$1.57 & 0.02 & 0.04 & 0.014 & 18.37 & 
0.03 & 47.1 & 0.6\\
IC 2146  & 17.01 & 0.02 & 72  & $-$1.56 & 0.02 & 0.12 & 0.041 & 18.53 & 
0.03 & 50.8 & 0.8\\
SL 663   & 16.84 & 0.04 & 29  & $-$1.52 & 0.02 & 0.04 & 0.014 & 18.35 & 
0.04 & 46.7 & 0.9\\
NGC 2121 & 16.83 & 0.02 & 184 & $-$1.51 & 0.02 & 0.10 & 0.034 & 18.31 & 
0.02 & 45.8 & 0.5\\
NGC 2173 & 16.94 & 0.03 & 62  & $-$1.53 & 0.04 & 0.10 & 0.034 & 18.44 & 
0.04 & 48.7 & 1.0\\
NGC 2155 & 16.78 & 0.02 & 63  & $-$1.53 & 0.02 & 0.03 & 0.010 & 18.30 & 
0.03 & 45.7 & 0.7\\
NGC 2162 & 17.10 & 0.03 & 72  & $-$1.49 & 0.18 & 0.03 & 0.010 & 18.58 & 
0.18 & 52.0 & 4.5\\
ESO 121  & 16.93 & 0.03 & 20  & $-$1.20 & 0.06 & 0.03 & 0.010 & 18.12 & 
0.06 & 42.1 & 1.3\\
NGC 2203 & 16.97 & 0.02 & 128 & $-$1.48 & 0.16 & 0.11 & 0.038 & 18.41 & 
0.17 & 48.1 & 3.8\\
NGC 2193 & 16.88 & 0.04 & 28  & $-$1.58 & 0.01 & 0.04 & 0.014 & 18.45 & 
0.04 & 48.9 & 0.9\\
SL 869   & 17.12 & 0.06 & 15  & $-$1.48 & 0.16 & 0.10 & 0.034 & 18.57 & 
0.17 & 51.7 & 4.3\\
SL 896   & 16.89 & 0.07 & 7   & $-$1.58 & 0.01 & 0.09 & 0.031 & 18.44 &  
0.07 & 48.7 & 1.6
\enddata
\tablecomments{All numbers are given in magnitudes unless otherwise 
noted.  }
\end{deluxetable}

\begin{deluxetable}{lccccccccccc}
\tabletypesize{\scriptsize}
\tablecaption{Effect of LMC Geometry\label{tab:geom}}
\tablewidth{0pt}
\tablehead{
\colhead{Geometry} &
\colhead{R.A.} & \colhead{Decl.} &
\colhead{$\Theta$} & \colhead{$i$} & \colhead{$\mmo$} & \colhead{$D_0$} 
\\
\colhead{(Reference)} & (J2000.0) & (J2000.0) & (deg) & (deg) & (mag) & 
(kpc)
}
\startdata
\citet{vdmcioni2001} & 5 29 00 & $-$69 30 00 &
122.5 $\pm$ 8.3 & 34.7 $\pm$ 6.2 & 18.40 $\pm$ 0.04 & 47.9 $\pm$ 0.9 \\
\citet{olsensalyk2002} & 5 19 38.0 & $-$69 27 05.2 &
145 $\pm$ 4 &35.8 $\pm$ 2.4 & 18.41 $\pm$ 0.04 & 48.1 $\pm$ 0.9 \\
\citet{vdmetal2002} & 5 27 36 & $-$69 52 12 &
129.9 $\pm$ 6.0 & 34.7 $\pm$ 6.2 & 18.40 $\pm$ 0.04 & 47.9 $\pm$ 0.9 \\
\citet{nikolaevetal2004} & 5 17 36 & $-$69 01 48 & 
151.0 $\pm$ 2.4 & 30.7 $\pm$ 1.1 & 18.41 $\pm$ 0.04 & 48.1 $\pm$ 0.9
\enddata
\tablecomments{Units of right ascension are in hours, minutes, and seconds
and units of declination are in degrees, arcminutes, and arcseconds.
Distances given are for the LMC center, calculated by combining our
cluster distances with the given LMC geometry.
} \end{deluxetable}

\begin{deluxetable}{lccccccccccc}
\tabletypesize{\scriptsize}
\tablecaption{LMC Center Distances\label{tab:cen_dist}}
\tablewidth{0pt}
\tablehead{
\colhead{Cluster} & 
\colhead{$D$} & \colhead{$\sigma_{D}$} &
\colhead{$D_0$} & \colhead{$\sigma_{D_0}$}
\\
\colhead{Name} & (mag) & (mag) & (mag) & (mag) 
}
\startdata
NGC 1651 & 18.46 & 0.03 & 18.35 & 0.04 \\
SL 61    & 18.49 & 0.09 & 18.30 & 0.10 \\
NGC 1783 & 18.02 & 0.18 & 18.05 & 0.18 \\
NGC 1846 & 18.13 & 0.19 & 18.14 & 0.19 \\
NGC 1978 & 18.40 & 0.02 & 18.47 & 0.03 \\
Hodge 4  & 18.37 & 0.03 & 18.47 & 0.04 \\
IC 2146  & 18.53 & 0.03 & 18.41 & 0.04 \\
SL 663   & 18.35 & 0.04 & 18.45 & 0.05 \\
NGC 2121 & 18.31 & 0.02 & 18.28 & 0.03 \\
NGC 2173 & 18.44 & 0.04 & 18.38 & 0.05 \\
NGC 2155 & 18.30 & 0.03 & 18.42 & 0.04 \\
NGC 2162 & 18.58 & 0.18 & 18.73 & 0.18 \\
ESO 121  & 18.12 & 0.06 & 18.33 & 0.08 \\
NGC 2203 & 18.41 & 0.17 & 18.29 & 0.17 \\
NGC 2193 & 18.45 & 0.04 & 18.58 & 0.05 \\
SL 869   & 18.57 & 0.17 & 18.60 & 0.17 \\
SL 896   & 18.44 & 0.07 & 18.49 & 0.07 
\enddata
\end{deluxetable}

\begin{deluxetable}{lccccccccccc}
\tabletypesize{\scriptsize}
\tablecaption{LMC Globular Cluster Information\label{tab:old_clust}}
\tablewidth{0pt}
\tablehead{
\colhead{Cluster} &
\colhead{R.A.} & \colhead{Decl.} &
\colhead{[Fe/H]} & \colhead{$V_{RR}$} & \colhead{$\ebv$} & 
\colhead{$D$} \\
(Name) & (J2000.0) & (J2000.0) & (dex) & (mag) & (mag) & (kpc)
}
\startdata
NGC 1466  & 03 44 33.35& $-$71 40 17.7& $-1.9 \pm 0.1$& 19.33 $\pm$ 0.02& 
0.05& 51.8 $\pm$ 1.0\\
Reticulum & 04 36 11.00& $-$58 51 40.0& $-1.7 \pm 0.1$& 19.07 $\pm$ 0.01& 
0.00& 47.3 $\pm$ 1.5\\
NGC 1841  & 04 45 23.83& $-$83 59 49.0& $-2.2 \pm 0.2$& 19.31 $\pm$ 0.02& 
0.11& 47.6 $\pm$ 1.9\\
NGC 1786  & 04 59 07.82& $-$67 44 42.8& $-2.3 \pm 0.2$& 19.27 $\pm$ 0.03& 
0.06& 50.8 $\pm$ 1.6\\
NGC 1835  & 05 05 06.58& $-$69 24 13.9& $-1.8 \pm 0.2$& 19.38 $\pm$ 0.05& 
0.09& 48.5 $\pm$ 2.0\\
NGC 2210  & 06 11 31.36& $-$69 07 17.0& $-1.9 \pm 0.2$& 19.12 $\pm$ 0.02& 
0.09& 43.5 $\pm$ 1.5\\
NGC 2257  & 06 30 13.00& $-$64 19 29.1& $-1.8 \pm 0.1$& 19.03 $\pm$ 0.02& 
0.04& 44.4 $\pm$ 0.8
\enddata
\tablecomments{Units of right ascension are in hours, minutes, and seconds
and units of declination are in degrees, arcminutes, and arcseconds.
} 
\end{deluxetable}


\end{document}